\documentclass[aps,prl,preprint]{revtex4-1}

\usepackage[pdftex]{graphicx}
\usepackage[pdftex,bookmarks=false,
                colorlinks,
                linkcolor=blue,
                citecolor=blue]{hyperref}
\usepackage{dcolumn}
\usepackage{bm}
\usepackage{braket,amsmath}
\usepackage{upgreek}
\usepackage{amsmath}
\usepackage{SIunits}
\usepackage{bbold}
\usepackage{multirow,varwidth}

\begin{document}


\title{Controlling spin relaxation with a cavity}

\author{A. Bienfait$^{1}$, J.J. Pla$^{2}$, Y. Kubo$^{1}$, X. Zhou$^{1,3}$, M. Stern$^{1,4}$, C.C. Lo$^{2}$, C.D. Weis$^{5}$, T. Schenkel$^{5}$, D. Vion$^{1}$, D. Esteve$^{1}$, J.J.L. Morton$^{2}$, and P. Bertet$^{1}$}

\affiliation{$^{1}$Quantronics group, Service de Physique de l'Etat Condens\'e, DSM/IRAMIS/SPEC, CNRS UMR 3680, CEA-Saclay,
91191 Gif-sur-Yvette cedex, France }

\affiliation{$^{2}$ London Centre for Nanotechnology, University College London, London WC1H 0AH, United Kingdom}
	
\affiliation{$^{3}$Institute of Electronics Microelectronics and Nanotechnology, CNRS UMR 8520, ISEN Department, Avenue Poincar\'e, CS 60069, 59652 Villeneuve d'Ascq Cedex, France}

\affiliation{$^{4}$ Quantum Nanoelectronics Laboratory, BINA, Bar Ilan University, Ramat Gan, Israel}

\affiliation{$^{5}$Accelerator Technology and Applied Physics Division, Lawrence Berkeley National Laboratory, Berkeley,
California 94720, USA}

\date{\today}

\maketitle

Spontaneous emission of radiation is one of the fundamental mechanisms by which an excited quantum system returns to equilibrium. For spins, however, spontaneous emission is generally negligible compared to other non-radiative relaxation processes because of the weak coupling between the magnetic dipole and the electromagnetic field. In 1946, Purcell realised~\cite{Purcell.PhysRev.69.681(1946)} that the spontaneous emission rate can be strongly enhanced  by placing the quantum system in a resonant cavity --- an effect which has since been used extensively to control the lifetime of atoms and semiconducting heterostructures coupled to microwave~\cite{Goy1983} or optical~\cite{Heinzen.PhysRevLett.58.1320(1987),Yamamoto1991337} cavities, underpinning single-photon sources~\cite{Gerard.PhysRevLett.81.1110(1998)}. Here we report the first application of these ideas to spins in solids. By coupling donor spins in silicon to a superconducting microwave cavity of high quality factor and small mode volume, we reach for the first time the regime where spontaneous emission constitutes the dominant spin relaxation mechanism. The relaxation rate is increased by three orders of magnitude when the spins are tuned to the cavity resonance, showing that energy relaxation can be engineered and controlled on-demand. Our results provide a novel and general way to initialise spin systems into their ground state, with applications in magnetic resonance and quantum information processing~\cite{Butler.PhysRevA.84.063407(2011)}.
They also demonstrate that, contrary to popular belief, the coupling between the magnetic dipole of a spin and the electromagnetic field can be enhanced up to the point where quantum fluctuations have a dramatic effect on the spin dynamics; as such our work represents an important step towards the coherent magnetic coupling of individual spins to microwave photons.  

\begin{figure}[htbp!]
\includegraphics[width=88mm]{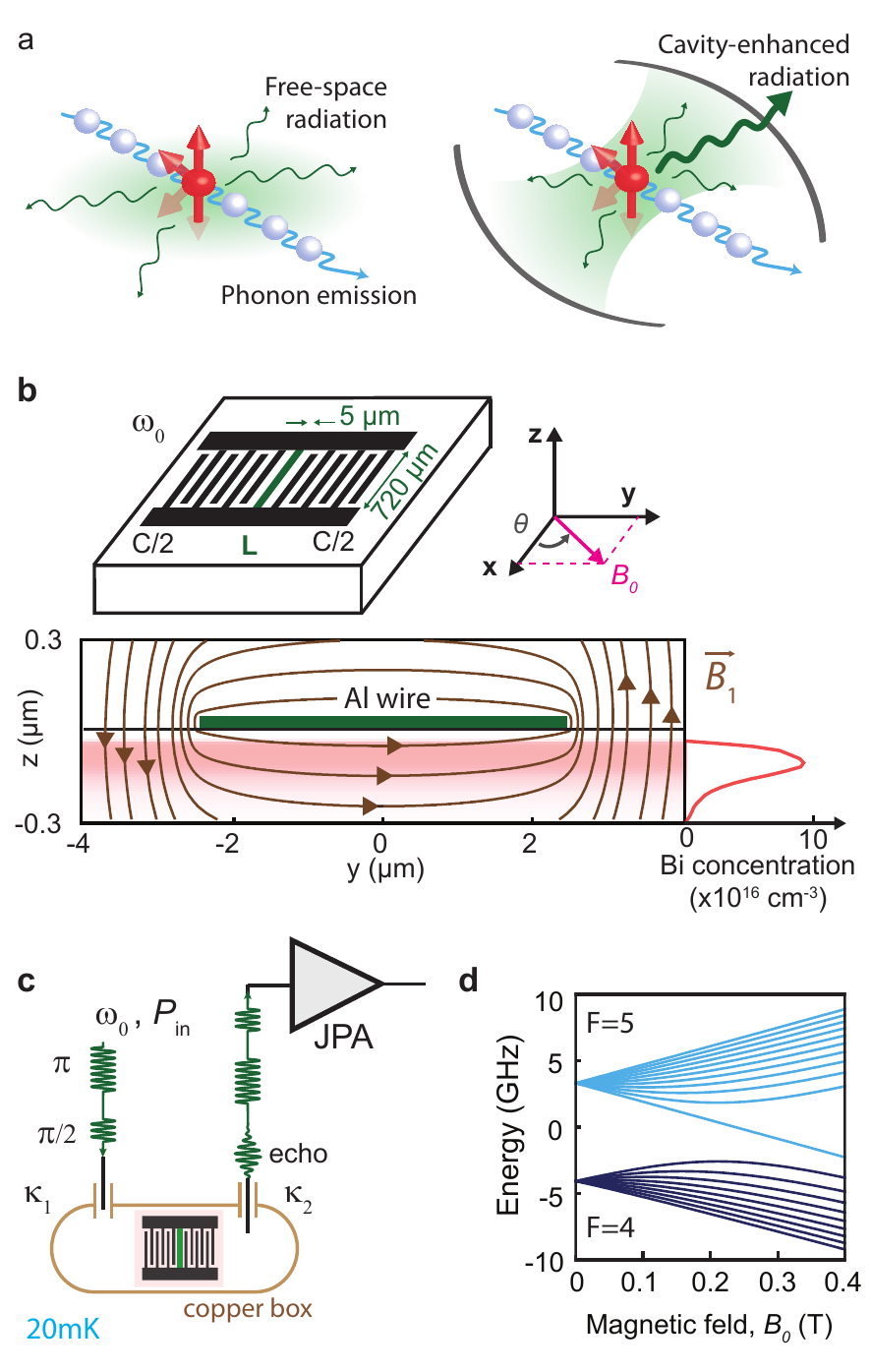}
\caption{\label{fig:figure1} 
\textbf{Purcell-enhanced spin relaxation and experimental setup.} 
\textbf{a)} By placing a spin in a resonant cavity, radiative spin relaxation can be made to dominate over intrinsic processes such as phonon-induced relaxation. 
\textbf{b)} (top) A planar superconducting resonator comprising an interdigitated capacitor in parallel with an inductive wire is fabricated on top of Bi-doped $^{28}$Si.
A static magnetic field $B_0$ is applied parallel to the ($x$-$y$) plane of the 50~nm thick aluminium layer with a tunable orientation given by $\theta$. (bottom) Magnetic field lines of the microwave excitation field $\vec{B_1}$ generated by the aluminium wire (arrows) are superimposed over the local concentration of Bi donors (red), obtained by secondary ion mass spectrometry (SIMS). 
\textbf{c)} The sample is mounted in a copper box thermally anchored at 20~mK and probed by microwave pulses via asymmetric antennae coupled with rate $\kappa_1 \approx \kappa_2 /5$ to the resonator. Microwave pulses at $\omega_0$ of power $P_{\rm in}$ are sent by antenna 1, and the microwave signal leaving via antenna 2 is directed to the input of a Josephson Parametric Amplifier (JPA).
\textbf{d)} Energy levels of the electron and nuclear spin of bismuth donors in silicon (see Suppl. Info.)}
\end{figure}

Spin relaxation is the process by which a spin reaches thermal equilibrium by exchanging an energy quantum $\hbar \omega_{\rm s}$ with its environment ($\omega_{\rm s}$ being its resonance frequency) for example in the form of a photon or a phonon, as shown in Fig.~\ref{fig:figure1}a. Understanding and controlling spin relaxation is of essential importance in applications such as spintronics~\cite{Sinova.NatureMat.11.368(2012)} and quantum information processing~\cite{Ladd.Nature.464.45(2010)} as well as magnetic resonance spectroscopy and imaging~\cite{Levitt.SpinDynamics}. For such applications, the spin relaxation time $T_1$ must be sufficiently long to permit coherent spin manipulation; however, if $T_1$ is too long it becomes a major bottleneck which limits the repetition rate of an experiment, and in turn impacts factors such as the achievable sensitivity.
Certain types of spins can be actively reset in their ground state by optical~\cite{Robledo.Nature.477.574(2011)} or electrical~\cite{Pla2012} means due to their specific energy level scheme, while methods such as chemical doping have been employed to influence spin relaxation times ex-situ~\cite{Shapiro.NatureBiotech.28.264(2010)}. Nevertheless, an efficient, general and tuneable initialization method for spin systems is still currently lacking.

At first inspection, spontaneous emission would appear an unlikely candidate to influence spin relaxation: for example, an electron spin in free space and at a typical frequency of $\omega_s / 2\pi \simeq 8$\,GHz, spontaneously emits a photon at a rate of $\sim 10^{-12}\,\mathrm{s}^{-1}$. 
However, the Purcell effect provides a means to dramatically enhance spontaneous emission, and thus gain precise and versatile control over spin relaxation~\cite{Purcell.PhysRev.69.681(1946)}.
Consider a spin embedded in a microwave cavity of quality factor $Q$ and frequency $\omega_0$. 
If the cavity damping rate $\kappa~= \omega_0 / Q$ is greater than the spin-cavity coupling $g$, the cavity then provides an additional channel for spontaneous emission of microwave photons, governed by a so-called Purcell rate~\cite{Butler.PhysRevA.84.063407(2011),Wood.PhysRevLett.112.050501(2014)}
\begin{equation}
\label{eq:Purcell}
\Gamma_{\rm P} = \kappa \frac{g^2} {\kappa^2 / 4 + \delta ^2},
\end{equation}
where $\delta~=\omega_0-\omega_{\rm s}$ is the spin-cavity detuning~(see Fig.~\ref{fig:figure1}a and Suppl. Info.). 

This cavity-enhanced spontaneous emission can be much larger than in free space, and is strongest when the spins and cavity are on-resonance ($\delta =0$), where $\Gamma_{\rm P} = 4 g^2 / \kappa $. Furthermore, the Purcell rate can be modulated by changing the coupling constant or the detuning, allowing spin relaxation to be tuned on-demand. The Purcell effect was used to detect spontaneous emission of radiofrequency radiation from nuclear spins coupled to a resonant circuit~\cite{Sleator.PhysRevLett.55.1742(1985)}, but even then the corresponding Purcell rate $\Gamma_{\rm P} \simeq 10^{-16}~\mathrm{s}^{-1}$ (or 1 photon emitted every 300 million years) was negligible compared to the intrinsic spin-lattice relaxation processes.     
In order for photon emission to become the dominant spin relaxation mechanism, both a large spin-cavity coupling and a low cavity damping rate are needed: in our experiment, this is achieved by combining the microwave confinement provided by a micron-scale resonator with the high quality factors enabled by the use of superconducting circuits.


The device consists of two planar aluminium lumped-element superconducting resonators patterned onto a silicon chip which was enriched in nuclear-spin-free $^{28}\mathrm{Si}$ and implanted with bismuth atoms (see Fig.~\ref{fig:figure1}b) at a sufficiently low concentration for collective radiation effects to be absent. A static magnetic field $\mathbf{B_0}$ is applied in the plane of the aluminium resonators, at an angle $\theta$ from the resonator inductive wire, tunable in-situ. The device is mounted inside a copper box and cooled to 20~mK. Each resonator can be used to perform inductive detection of the electron-spin resonance (ESR) signal of the bismuth donors: microwave pulses at $\omega_0$ are applied at the resonator input, generating an oscillating magnetic field $B_1$ around the inductive wire which drives the surrounding spins; the quantum fluctuations of this field, present even when no microwave is applied, are responsible for the Purcell spontaneous emission. Hahn echo pulse sequences~\cite{Hahn1950} are used, resulting in the emission of a spin-echo in the detection waveguide, which is amplified with a sensitivity reaching the quantum limit thanks to the use of a Josephson Parametric Amplifier~\cite{Zhou.PhysRevB.89.214517(2014)} and demodulated, yielding the integrated echo signal quadrature $A_{\rm Q}$. A more detailed setup description can be found in~\cite{QuantumESR}.

\begin{figure}[htbp!]
\includegraphics[width=88mm]{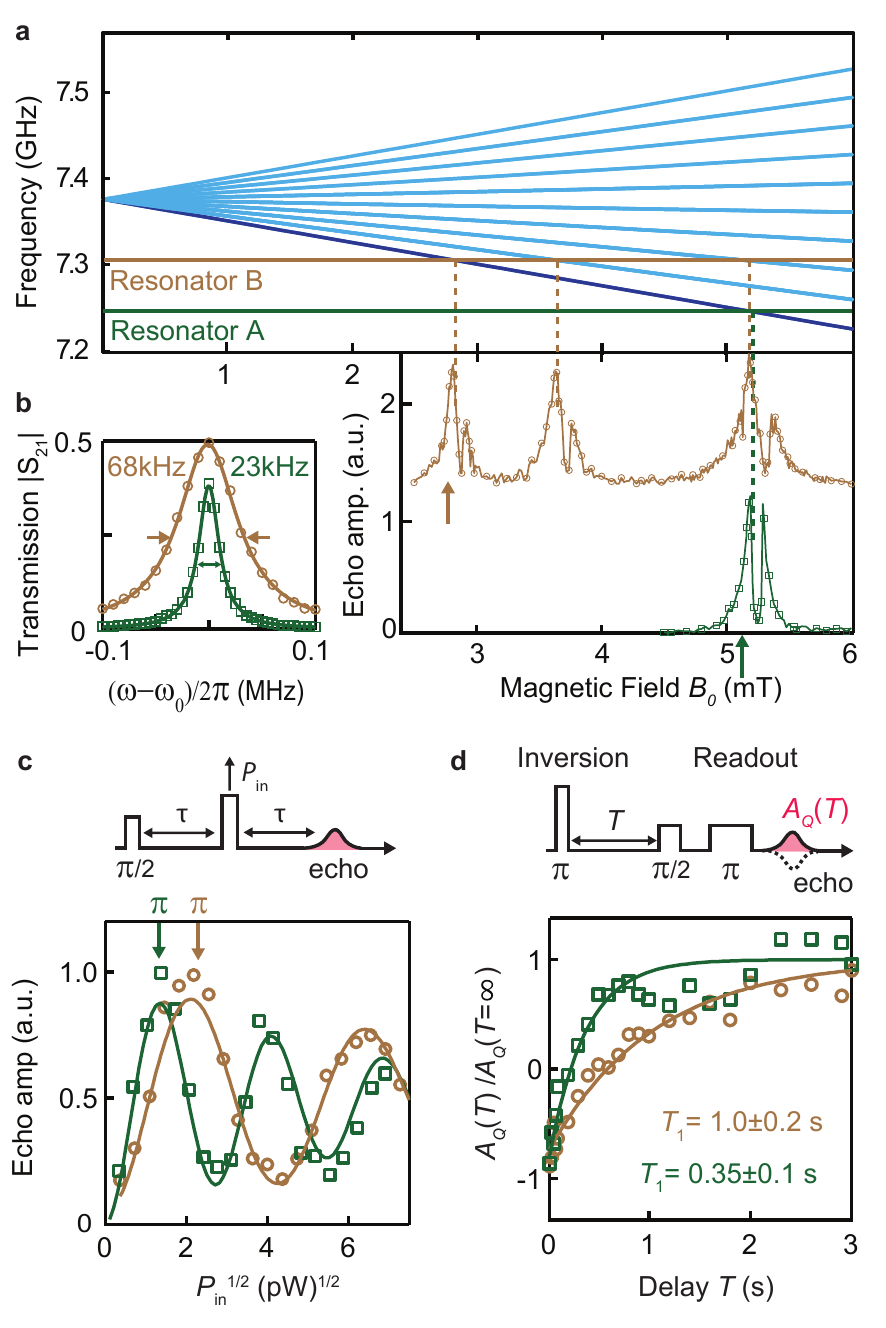}
\caption{\label{fig:figure2}
\footnotesize{
\textbf{ESR spectroscopy and Purcell-limited $T_1$ measurement.}
\textbf{a)} (top) Dominant electron spin resonance transitions of the Si:$^{209}$Bi spin system (see Suppl. Info). 
We employ two resonators, A and B, with frequencies 7.246 and $7.305$~GHz, which cross up to three spin transitions in the magnetic field range 0 -- 6~mT, as seen in the echo-detected magnetic field sweep (bottom panel, vertically offset for clarity).
Subsequent spin relaxation measurements were made at the magnetic fields indicated by the arrows, corresponding to the $\ket{F,m_F} = \ket{4,-4}\leftrightarrow\ket{5,-5}$ transition for each resonator. The doublet structure of each transition is caused by strain exerted by the aluminum film on the donors (see Suppl. Info).
\textbf{b)} Cavity linewidths for resonators A and B are measured to be 23 and 68~kHz respectively.
\textbf{c)} Rabi oscillations are driven by varying the cavity input power of the refocusing $5$-$\upmu$s-long  $\pi$ pulse. 
\textbf{d)} The inversion-recovery sequence is used to measure the spin relaxation time $T_1$. Spin polarisation is measured with a Hahn echo sequence (50-$\upmu$s-long $\pi/2$ pulse, delay $\tau = 500$~$\upmu$s, and 100-$\upmu$s-long $\pi$ pulse). Rescaled by its value for $T \gg T_1$, $A_Q$ goes from $-1$ when the spins are fully inverted to $+1$ at thermal equilibrium. The pulse durations were chosen such that only spins within a narrow spectral range were detected, producing a well-defined Purcell-limited $T_1$ (see Suppl.\ Info). Data in this figure were obtained with the static field $B_0$ parallel to the inductor ($\theta = 0$).}
}
\end{figure}

Bismuth is a donor in silicon~\cite{Feher.PhysRev.114.1219(1959)} with a nuclear spin $I = 9/2$. At cryogenic temperatures it can bind an electron (with spin $S = 1/2$) in addition to those shared with the surrounding Si lattice. The large hyperfine interaction $A\overrightarrow{S}\cdot\overrightarrow{I}$ between the electron and nuclear spin, where $\overrightarrow{S}$ and $\overrightarrow{I}$ are the electron and nuclear spin operators and $A/h = 1475$~MHz, produces a splitting of 7.375~GHz between the ground and excited multiplets at zero magnetic field (see Fig.~\ref{fig:figure1}d for the complete energy diagram~\cite{Wolfowicz.PhysRevB.86.245301(2012)}). This makes the system ideal for coupling to superconducting circuits~\cite{Morley.NatureMat.9.725(2010),George.PhysRevLett.105.067601(2010)}. At low fields ($B_0 < 10$\,mT, compatible with the critical field of aluminum) all $\Delta m_{\rm F} = \pm 1$ transitions are allowed, $m_{\rm F}$ being the projection of the total spin ($\overrightarrow{F} = \overrightarrow{I} + \overrightarrow{S}$) along $B_0$. Considering only the transitions with largest matrix element, resonator A ($\omega_{0A}/2\pi = 7.245$~GHz, $Q_A=3.2 \times 10^5$) crosses the $\ket{F,m_F}=\ket{4,-4} \leftrightarrow \ket{5,-5}$ transition, whilst resonator B ($\omega_{0B}/2\pi = 7.305$~GHz, $Q_B=1.1 \times 10^5$) crosses the transitions $\ket{4,-4} \leftrightarrow \ket{5,-5}$, $\ket{4,-3} \leftrightarrow \ket{5,-4}$, and $\ket{4,-2} \leftrightarrow \ket{5,-3}$ (see Figs.~\ref{fig:figure2}a and b).

The echo signal $A_{\rm Q}$ from each resonator as a function of $B_0$ shows resonances at the expected magnetic fields, split into two peaks of full-width-half-maximum $\Delta \omega / 2\pi \sim 2$~MHz (see Fig.~\ref{fig:figure2}a). As explained in~\cite{QuantumESR}, this splitting is believed to be the result of strain induced in the silicon by the aluminium surface structure, which is non-negligible at the donor implant depth of $\sim 100$~nm. In the following we focus on the lower-frequency peak of the $\ket{4,-4} \leftrightarrow \ket{5,-5}$ line which corresponds to spins lying under the wire (see Suppl. Info). Over the region occupied by these spins, the $B_1$ field amplitude varies by less than $\pm 2\%$, as evidenced by the well-defined Rabi oscillations observed when we sweep the power of the refocusing pulse $P_{\rm in}$ at the cavity input (see Fig.~\ref{fig:figure2}c), allowing us to determine the input power of a $\pi$ pulse for a given pulse duration. 

We measure the relaxation time $T_1$ by performing an ``inversion-recovery'' experiment~\cite{SchweigerEPR(2001)} (see schematic, top of Fig.~\ref{fig:figure2}a), with the static field $B_0$ aligned along $x$ ($\theta = 0$). A $\pi$ pulse first inverts the spins whose frequency lies within the resonator bandwidth $\kappa_A/ 2\pi = 23$\,kHz (or $\kappa_B/ 2\pi = 68$\,kHz); note that this constitutes a small subset of the total number of spins since $\kappa_{A,B} \ll \Delta \omega$. After a varying delay $T$, a Hahn echo sequence provides a measure of the longitudinal spin polarization. Fitting the data with decaying exponentials, we extract $T_1 = 0.35$\,s for resonator A and $T_1 = 1.0$\,s for resonator B.


For a quantitative comparison with the expected Purcell rate, it is necessary to evaluate the spin-resonator coupling constant $g = \gamma_e \langle F,m_F | S_x | F+1,m_{F}-1 \rangle \left\| \mathbf{\delta B_\bot} \right\|$, where $\gamma_{\rm e} / 2\pi \simeq 28$\,GHz/T is the electronic gyromagnetic ratio and $\mathbf{\delta B_\bot} $ is the component of the resonator field vacuum fluctuations orthogonal to $\mathbf{B_0}$ (see Suppl. Info of \cite{QuantumESR}). A numerical estimate yields $g_0 / 2\pi = 56 \pm 1$\,Hz for the spins located below the resonator inductive wire that are probed in our measurements and for $\theta = 0$. An independent estimate is obtained by measuring Rabi oscillations. Their frequency $\Omega_R = 2 g_0 \sqrt{\bar{n}}$ directly yields $g_0$ upon knowledge of the average intra-cavity photon number $\bar{n}$, which can be determined with a $\sim 30 \%$ imprecision from $P_{\rm in}$ and the measured resonator coupling to the input and output antennae (see Suppl. Info). We obtain $g_0 / 2\pi = 50 \pm 7$\,Hz for resonator A and $58 \pm 7$\,Hz for resonator B, compatible with the numerical estimate. The corresponding resonant Purcell spontaneous emission time is $\Gamma_P^{-1} = 0.36 \pm 0.09$\,s for resonator A and $0.81 \pm 0.17$\,s for resonator B, in agreement with the experimental values.

\begin{figure}[htbp!]
\includegraphics[width=88mm]{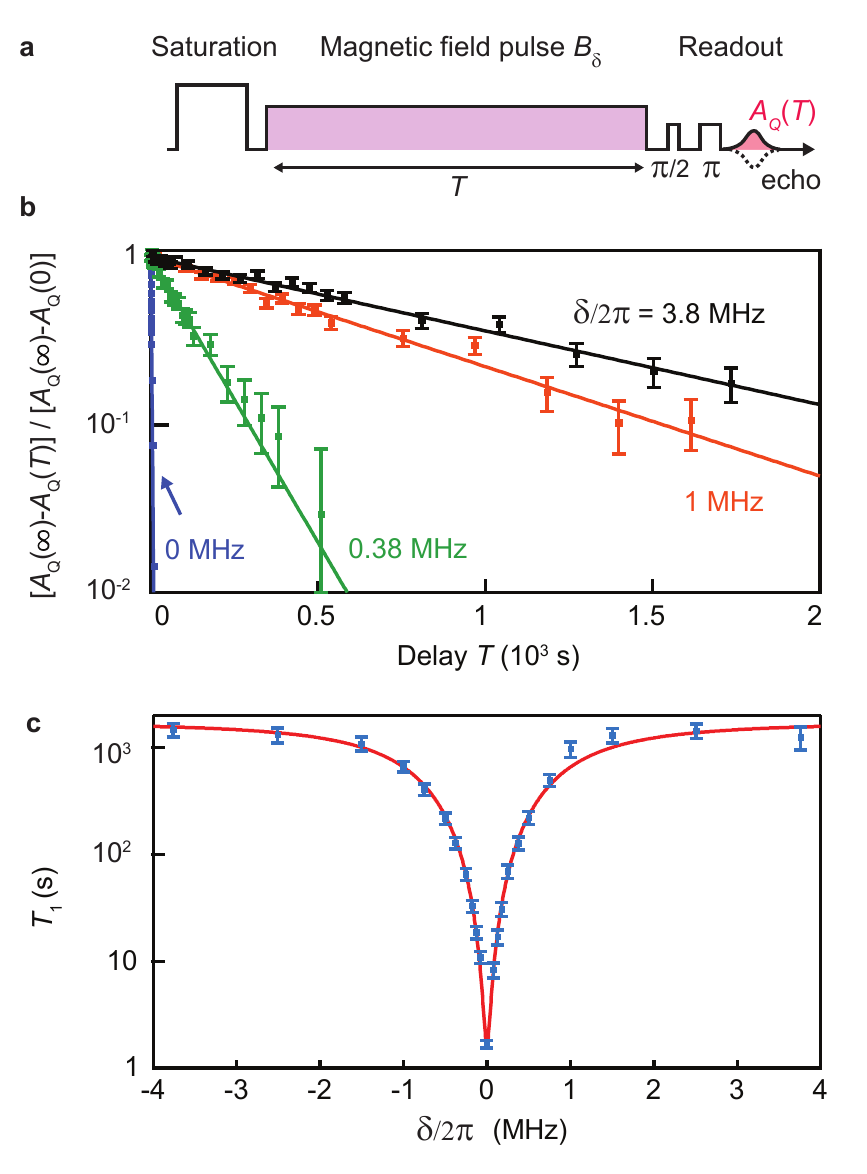}
\caption{\label{fig:figure4} 
\textbf{Controlling Purcell relaxation by spin-cavity detuning.}
\textbf{a)} In-between their saturation and subsequent readout, the spins are detuned from the cavity by $\delta=\frac{d \omega_{\rm s}}{dB} B_\delta$ by applying a magnetic field pulse of amplitude $B_\delta$, with $\frac{1}{2\pi}\frac{d \omega_{\rm s}}{dB} \simeq 25$~GHz/T for this transition and magnetic field. 
\textbf{b)} The decay of spin polarisation is well fit (lines) to exponential decays, with relaxation time constants $T_1$ increasing with the detuning.  
\textbf{c)} Measured $T_1$ as a function of detuning $\delta$. Line is a fit with $(\Gamma_{\rm P}(\delta) + \Gamma_{\rm NR})^{-1}$, yielding $\Gamma_{\rm NR}^{-1}=1600$\,s. Note that these measurements are taken using resonator B and with $\theta = \pi /4$, which results in $T_1=1.7$\,s at $\delta = 0$. The angle $\theta$ changes by at most $10\%$ during the magnetic field pulse.}
\end{figure}

According to Eq.~\ref{eq:Purcell}, a Purcell-limited $T_1$ should be strongly dependent on the spin-cavity detuning. We introduce a magnetic field pulse of duration $T$ between the spin excitation and the spin-echo sequence (see Fig.~\ref{fig:figure4}a), which results in a temporary detuning $\delta$ of the spins. The echo signal amplitude $A_{\rm Q}$ as a function of $T$ yields their energy relaxation time while they are detuned by $\delta$. To minimize the influence of spin diffusion~\cite{SchweigerEPR(2001)}, the spin excitation is performed here by a high-power long-duration saturating pulse (see Fig.~\ref{fig:figure4}a and Suppl. Info) instead of an inversion pulse as in Fig.~\ref{fig:figure2}d. As evident in Fig.~\ref{fig:figure4}b, we find that the decay of the echo signal is well fit by a single exponential with a decay time increasing with $|\delta|$. The extracted $T_1(\delta)$ curve (see Fig.~\ref{fig:figure4}c) shows a remarkable increase of $T_1$ by up to $3$ orders of magnitude when the spins are detuned away from resonance, until it becomes limited by a non-radiative energy decay mechanism with rate $\Gamma_{\rm NR}^{-1} = 1600 \pm 300$\,s. Given the doping concentration in our sample, this rate is consistent with earlier measurements of donor spin relaxation times~\cite{Feher.PhysRev.114.1245(1959)}, which have been attributed to charge hopping, but could also arise here from spatial diffusion of the spin magnetisation away from the resonator mode volume. Figure ~\ref{fig:figure4}c shows that the $T_1(\delta)$ measurements are in agreement with the expected dependence $(\Gamma_{\rm P}(\delta) + \Gamma_{\rm NR})^{-1}$, with the only free parameter in this fit being $\Gamma_{\rm NR}$. 


\begin{figure}[htbp!]
\includegraphics[width=88mm]{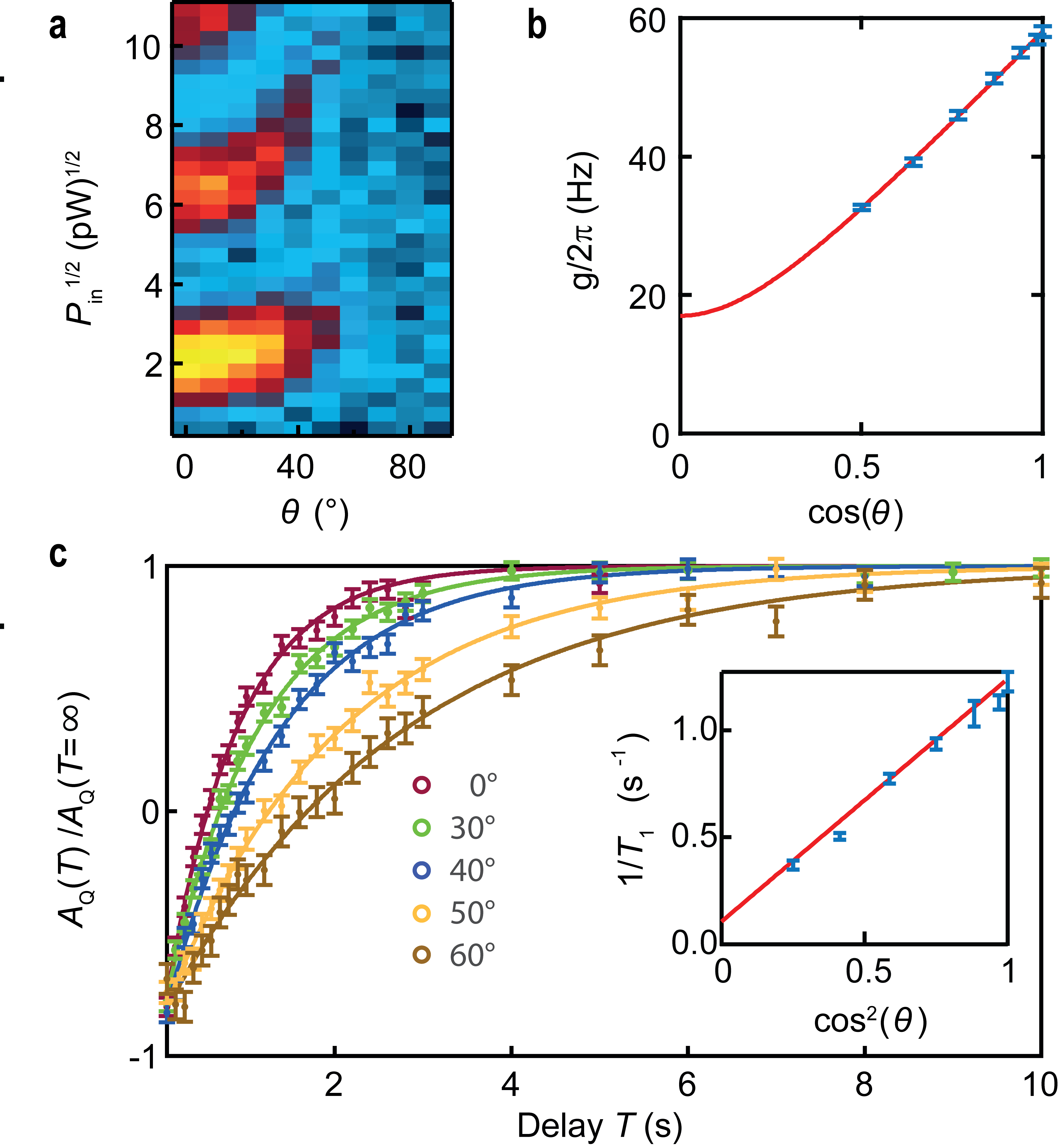}
\caption{\label{fig:figure3} 
\textbf{Dependence of Purcell relaxation on spin-cavity coupling $g$.}
\textbf{a)}  Rabi oscillations (as in Fig.~\ref{fig:figure2}c) measured as a function of field orientation $\theta$ (see Fig.\ref{fig:figure1}b) are used to extract \textbf{b)} the spin-cavity coupling strength $g$. Its dependence on $\theta$ is fit to the expression in the main text (red line); the non-zero value of $g(\pi/2)$ is due to the finite out-of-plane component of the microwave magnetic field.
\textbf{c)} Inversion-recovery measurements as a function of $\theta$ confirm that the relaxation time $T_1$ (see inset) varies as $g(\theta)^2$. Solid line is the Purcell formula prediction using the $g(\theta)$ dependence fitted from (b). All data in the figure were taken using resonator B.}
\end{figure}
Having demonstrated the effect of cavity linewidth and detuning on the Purcell rate, we finally explore the effect of modulating the spin-cavity coupling constant $g$. This can be achieved by varying the orientation $\theta$ of the static magnetic field $B_0$ in the $x$-$y$ plane (Fig.~\ref{fig:figure1}b), adjusting the component of the microwave magnetic field (mostly along $y$ under the inductive wire) which is orthogonal to $B_0$. More precisely, $g(\theta) = \gamma_{\rm e} \langle F,m_F | S_x | F+1,m_{F}-1 \rangle \sqrt{\delta B_{1y}^2 \cos(\theta)^2 + \delta B_{1z}^2}$ (noting that $\delta B_{1x}=0$), and we expect $\delta B_{1z}\ll\delta B_{1y}$ for the spins lying under the wire that are probed in these measurements.
This is verified experimentally by measuring the Rabi frequency as a function of $\theta$, as shown in Fig.~\ref{fig:figure3}a \& b, allowing us to extract 
$g(0)/2\pi = 58$\,Hz and 
$g(\pi/2)/2\pi = 17$\,Hz. As expected, we measure longer spin relaxation times for increasing values of $\theta$, as shown in Fig.~\ref{fig:figure3}c, with the relaxation rate $T_1^{-1}$ scaling as $g^2 (\theta)$, in agreement with Eq.~\ref{eq:Purcell}.
 Overall, the data of Figs.~\ref{fig:figure4} and \ref{fig:figure3} demonstrate unambiguously that cavity-enhanced spontaneous emission is by far the dominant spin relaxation channel when the spins are resonant with the cavity, since the probability for a spin-flip to occur due to emission of a microwave photon in the cavity is $1/[1+\Gamma_{\rm NR}/\Gamma_{\rm P}(\delta = 0)] = 0.999$, very close to unity.


At this point it is interesting to reflect on the important fact that the spontaneous emission evidenced here is an energy relaxation mechanism which does not require the presence of a macroscopic magnetization to be effective. Under the Purcell effect, each spin independently relaxes towards thermal equilibrium by microwave photon emission, so that the sample ends up in a fully polarized state after a time longer than $\Gamma_{\rm P}^{-1}$, regardless of its initial state. This is in stark contrast with the well-known phenomenon of radiative damping~\cite{Bloembergen1954} of a transverse magnetization generated by earlier microwave pulses, which is a coherent collective effect under which the degree of polarization of a sample cannot increase. We also note that had our device possessed a larger spin concentration, spontaneous relaxation would have occurred collectively, manifesting itself as a non-exponential decay of the echo signal on a time scale faster than $\Gamma_P^{-1}$~\cite{Wood.PhysRevLett.112.050501(2014)} and leading to an incomplete thermalization~\cite{Butler.PhysRevA.84.063407(2011),Wood.Arxiv.1506.03007}. Such superradiant or maser emission~\cite{Feher.PhysRev.109.221(1958)} requires the dimensionless parameter $C = N g^2 / (\kappa \Delta \omega)$ called cooperativity ($N$ being the total number of spins) to satisfy $C \gg 1$~\cite{Butler.PhysRevA.84.063407(2011),Temnov.PhysRevLett.95.243602(2005),Wood.Arxiv.1506.03007}, which is not the case here because of the large inhomogeneous broadening of the spin resonance caused by strain.



Our demonstrated ability to modulate spin relaxation through 3 orders of magnitude by changing the applied field by less than 0.1~mT opens up new perspectives for spin-based quantum information processing: long intrinsic relaxation times which are desirable to maximise the spin coherence time can be combined with fast, on-demand initialisation of the spin state. 
We also anticipate Purcell relaxation will offer a powerful approach to dynamical nuclear polarisation~\cite{Carver.PhysRev.92.212.2(1953),Abragam.RepProgrPhys.41.395(1978)}, for example by tuning the cavity to match an electron-nuclear spin flip-flop transition, enhancing the rate of cross-relaxation to pump polarisation into the desired nuclear spin state~\cite{Bloembergen.PhysRev.114.445(1959)}. 
%
%
The Purcell rate we obtain could be increased by reducing the transverse dimensions of the inductor wire to yield larger coupling constants, up to $5-10$\,kHz, bringing the spontaneous emission time below $1$\,ms (enabling faster repetition rates), as well as a higher sensitivity~\cite{QuantumESR}, up to the single-spin detection limit. 
Finally, our measurements constitute the first evidence that vacuum fluctuations of the microwave field can affect the dynamics of spins, and are thus a step towards the application of circuit quantum electrodynamics concepts to individual spins in solids.


\begin{acknowledgments}
We acknowledge technical support from P. S{\'e}nat, D. Duet, J.-C. Tack, P. Pari, P. Forget, as well as useful discussions within the Quantronics group. We acknowledge support of the European Research Council under the European Community's Seventh Framework Programme (FP7/2007-2013) through grant agreements No. 615767 (CIRQUSS), 279781 (ASCENT), and 630070 (quRAM), and of the C'Nano IdF project QUANTROCRYO. J.J.L.M. is supported by the Royal Society. C.C.L. is supported by the Royal Commission for the Exhibition of 1851. T.S. and C.D.W. were supported by the U. S. Department of Energy under contract DE-AC02-05CH11231.
\end{acknowledgments}

\hrulefill
\clearpage
\clearpage
\newpage
\widetext
\begin{center}
\textbf{\large Supplementary Material: Controlling spin relaxation with a cavity}
\end{center}
\setcounter{equation}{0}
\setcounter{figure}{0}
\setcounter{table}{0}
\setcounter{page}{1}
\makeatletter
\renewcommand{\theequation}{S\arabic{equation}}
\renewcommand{\thefigure}{S\arabic{figure}}

\newcommand{\ac}[0]{\ensuremath{\hat{a}}}
\newcommand{\adagc}[0]{\ensuremath{\hat{a}^{\dagger}}}
\newcommand{\ain}[0]{\ensuremath{\hat{a}_{\mathrm{in}}}}
\newcommand{\aout}[0]{\ensuremath{\hat{a}_{\mathrm{out}}}}
\newcommand{\aR}[0]{\ensuremath{\hat{a}_{\mathrm{R}}}}
\newcommand{\aT}[0]{\ensuremath{\hat{a}_{\mathrm{T}}}}
\renewcommand{\b}[0]{\ensuremath{\hat{b}}}
\newcommand{\bdag}[0]{\ensuremath{\hat{b}^{\dagger}}}
\newcommand{\betaI}[0]{\ensuremath{\beta_\mathrm{I}}}
\newcommand{\betaR}[0]{\ensuremath{\beta_\mathrm{R}}}
\newcommand{\bid}[0]{\ensuremath{\hat{b}_{\mathrm{id}}}}
\renewcommand{\c}[0]{\ensuremath{\hat{c}}}
\newcommand{\cdag}[0]{\ensuremath{\hat{c}^{\dagger}}}
\newcommand{\CorrMat}[0]{\ensuremath{\boldsymbol\gamma}}
\newcommand{\Deltacs}[0]{\ensuremath{\Delta_{\mathrm{cs}}}}
\newcommand{\Deltacsmax}[0]{\ensuremath{\Delta_{\mathrm{cs}}^{\mathrm{max}}}}
\newcommand{\Deltacsparked}[0]{\ensuremath{\Delta_{\mathrm{cs}}^{\mathrm{p}}}}
\newcommand{\Deltacstarget}[0]{\ensuremath{\Delta_{\mathrm{cs}}^{\mathrm{t}}}}
\newcommand{\Deltae}[0]{\ensuremath{\Delta_{\mathrm{e}}}}
\newcommand{\Deltahfs}[0]{\ensuremath{\Delta_{\mathrm{hfs}}}}
\newcommand{\dens}[0]{\ensuremath{\hat{\rho}}}
\newcommand{\e}[1]{\ensuremath{\times 10^{#1}}}
\newcommand{\erfc}[0]{\ensuremath{\mathrm{erfc}}}
\newcommand{\Fq}[0]{\ensuremath{F_{\mathrm{q}}}}
\newcommand{\gammapar}[0]{\ensuremath{\gamma_{\parallel}}}
\newcommand{\gammaperp}[0]{\ensuremath{\gamma_{\perp}}}
\newcommand{\gavg}[0]{\ensuremath{\mathcal{G}_{\mathrm{avg}}}}
\newcommand{\gbar}[0]{\ensuremath{\bar{g}}}
\newcommand{\gens}[0]{\ensuremath{g_{\mathrm{ens}}}}
\renewcommand{\H}[0]{\ensuremath{\hat{H}}}
\renewcommand{\Im}[0]{\ensuremath{\mathrm{Im}}}
\newcommand{\kappac}[0]{\ensuremath{\kappa_{\mathrm{c}}}}
\newcommand{\kappaL}[0]{\ensuremath{\kappa_{\mathrm{L}}}}
\newcommand{\kappamin}[0]{\ensuremath{\kappa_{\mathrm{min}}}}
\newcommand{\kappamax}[0]{\ensuremath{\kappa_{\mathrm{max}}}}
\newcommand{\kB}[0]{\ensuremath{k_{\mathrm{B}}}}
\newcommand{\mat}[1]{\ensuremath{\mathbf{#1}}}
\newcommand{\mean}[1]{\ensuremath{\langle#1\rangle}}
\newcommand{\namp}[0]{\ensuremath{n_{\mathrm{amp}}}}
\renewcommand{\neq}[0]{\ensuremath{n_{\mathrm{eq}}}}
\newcommand{\Nmin}[0]{\ensuremath{N_{\mathrm{min}}}}
\newcommand{\nsp}[0]{\ensuremath{n_{\mathrm{sp}}}}
\newcommand{\omegac}[0]{\ensuremath{\omega_{\mathrm{c}}}}
\newcommand{\omegas}[0]{\ensuremath{\omega_{\mathrm{s}}}}
\newcommand{\pauli}[0]{\ensuremath{\hat{\sigma}}}
\newcommand{\pexc}[0]{\ensuremath{p_{\mathrm{exc}}}}
\newcommand{\pexceff}[0]{\ensuremath{p_{\mathrm{exc}}^{\mathrm{eff}}}}
\newcommand{\Pa}[0]{\ensuremath{\hat{P}_{\mathrm{c}}}}
\newcommand{\Qmin}[0]{\ensuremath{Q_{\mathrm{min}}}}
\newcommand{\Qmax}[0]{\ensuremath{Q_{\mathrm{max}}}}
\renewcommand{\Re}[0]{\ensuremath{\mathrm{Re}}}
\renewcommand{\S}[0]{\ensuremath{\hat{S}}}
\newcommand{\Sminuseff}[0]{\ensuremath{\hat{S}_-^{\mathrm{eff}}}}
\newcommand{\Sxeff}[0]{\ensuremath{\hat{S}_x^{\mathrm{eff}}}}
\newcommand{\Syeff}[0]{\ensuremath{\hat{S}_y^{\mathrm{eff}}}}
\newcommand{\tildeac}[0]{\ensuremath{\tilde{a}_{\mathrm{c}}}}
\newcommand{\tildepauli}[0]{\ensuremath{\tilde{\sigma}}}
\newcommand{\Tcaveff}[0]{\ensuremath{T_{\mathrm{cav}}^{\mathrm{eff}}}}
\newcommand{\Techo}[0]{\ensuremath{T_{\mathrm{echo}}}}
\newcommand{\Tmem}[0]{\ensuremath{T_{\mathrm{mem}}}}
\newcommand{\Tswap}[0]{\ensuremath{T_{\mathrm{swap}}}}
\newcommand{\Var}[0]{\ensuremath{\mathrm{Var}}}
\renewcommand{\vec}[1]{\ensuremath{\mathbf{#1}}}
\newcommand{\Xa}[0]{\ensuremath{\hat{X}_{\mathrm{c}}}}
\newcommand{\Xid}[0]{\ensuremath{\hat{X}_{\mathrm{id}}}}
\newcommand{\Xin}[0]{\ensuremath{\hat{X}_{\mathrm{in}}}}
\newcommand{\Xout}[0]{\ensuremath{\hat{X}_{\mathrm{out}}}}
\newcommand{\Yin}[0]{\ensuremath{\hat{Y}_{\mathrm{in}}}}
\newcommand{\Yout}[0]{\ensuremath{\hat{Y}_{\mathrm{out}}}}

\section{Bismuth donor spin}

The spins used in this experiment, neutral bismuth donors in silicon, have  a nuclear spin $I=9/2$  and an electron spin $S=1/2$. The Halmitonian describing the system\cite{Wolfowicz.NatureNano.8.561(2013)} is 

\begin{equation}
  \H /\hbar = \vec{B} \cdot (\gamma_e \vec{S} \otimes \mathbb{1} - \gamma_n \mathbb{1} \otimes \vec{I}) + A\, \vec{S} \cdot \vec{I},
\label{eq:Hamiltonian_BiSI}
\end{equation}

where $A/h= 1.45\,$ GHz is the hyperfine coupling between electron and nuclear spins, $\gamma_e/ 2\pi=27.997$ GHz/T and $\gamma_n/ 2\pi=6.9$ MHz/T are the electronic and nuclear gyromagnetic ratios. In the limit of a small static magnetic field ($B_0 \lesssim 50\,$mT), the 20 electro-nuclear energy states are well approximated by eigenstates of the total angular momentum $\vec{F}=\vec{S}+\vec{I}$ and its projection, $m_F$, along the axis of the applied field. These eigenstates can be grouped in an $F=4$ ground and an $F=5$ excited multiplet separated by a frequency of $(I+1/2)A/h=7.35\,$GHz in zero-field, as shown in Figure 1 of the main text. 

For a given weak static field $B_0$ oriented along $\vec{z}$, transitions verifying $\Delta F\Delta m_F=\pm 1$ may be probed with an excitation field orientated along $\vec{x}$ (or $\vec{y}$) since their associated matrix element $\bra{F,m_F} S_x \ket{F+1,m_F\pm 1}=\bra{F,m_F} S_y \ket{F+1,m_F\pm 1}$ has the same magnitude as an ideal electronic spin 1/2 transition $\bra{m_s} S_y \ket{m_{s'}}=0.5$. Characteristics for the nine $\Delta F \Delta m_F=+1$ transitions and the nine $\Delta F \Delta m_F=-1$ transitions are given in Table~\ref{tbl:Transitions} for $B_0=3\,$mT. Only the ten transitions with a matrix element greater than 0.25 are shown on Figure 2a of the main text. Note that the eight hidden transitions with lower matrix element are actually degenerate with transitions with stronger matrix element. The transitions probed by our resonators are highlighted in red in Table~\ref{tbl:Transitions}.
\clearpage
\newcommand\NameEntry[1]{%
  \multirow{3}*{%
    \begin{minipage}{5em}
    #1%
    \end{minipage}}}
\newcolumntype{.}{D{.}{.}{0}}
\begin{table}[!h]
\setlength{\tabcolsep}{5pt}
\centering
\begin{tabular}{|p{4cm} .|p{4cm} .|r|.|}
\hline
\multicolumn{2}{|c|}{Transitions $\Delta F\Delta m_F=-1$} & \multicolumn{2}{c|}{Transitions $\Delta F\Delta m_F=+1$} & \NameEntry{Frequency \\ (GHz)} &  \NameEntry{$df/dB$ \\ (GHz/T)} \\ 
\multicolumn{2}{|l|}{{\scriptsize $\ket{F,m_F} \leftrightarrow \ket{F+1,m_F-1}$}} & \multicolumn{2}{l|}{{\scriptsize$\ket{F,m_F} \leftrightarrow \ket{F+1,m_F+1}$}}  &  &  \\ 
\multicolumn{2}{|r|}{{\scriptsize$\bra{F,m_F} S_x \ket{F+1,m_F-1}$}} & \multicolumn{2}{r|}{{\scriptsize$\bra{F,m_F} S_x \ket{F+1,m_F+1}$}}  & &  \\ 
\hline 
\textcolor{red}{$\ket{4,-4} \leftrightarrow \ket{5,-5}$} & $0.474$ &  & & 7.300 & $-25.1$ \\
\textcolor{red}{$\ket{4,-3} \leftrightarrow \ket{5,-4}$} & $0.423$ & $\ket{4,-4} \leftrightarrow \ket{5,-3}$ & $0.072$ & 7.317 & $-19.2$ \\
\textcolor{red}{$\ket{4,-2} \leftrightarrow \ket{5,-3}$} & $0.372$ & $\ket{4,-3} \leftrightarrow \ket{5,-2}$ & $0.125$ & 7.334 & $-13.8$ \\
$\ket{4,-1} \leftrightarrow \ket{5,-2}$ & $0.321$ & $\ket{4,-2} \leftrightarrow \ket{5,-1}$ & $0.176$ & 7.351 & $-8.1$ \\
$\ket{4, 0} \leftrightarrow \ket{5,-1}$ & $0.271$ & $\ket{4,-1} \leftrightarrow \ket{5, 0}$ & $0.226$ & 7.368 & $-2.5$ \\
$\ket{4, 1} \leftrightarrow \ket{5, 0}$ & $0.221$ & $\ket{4, 0} \leftrightarrow \ket{5, 1}$ & $0.277$ & 7.385 & $3.1$ \\
$\ket{4, 2} \leftrightarrow \ket{5, 1}$ & $0.171$ & $\ket{4, 1} \leftrightarrow \ket{5, 2}$ & $0.327$ & 7.401 & $8.7$ \\
$\ket{4, 3} \leftrightarrow \ket{5, 2}$ & $0.120$ & $\ket{4, 2} \leftrightarrow \ket{5, 3}$ & $0.376$ & 7.418 & $14.2$ \\
$\ket{4, 4} \leftrightarrow \ket{5, 3}$ & $0.069$ & $\ket{4, 3} \leftrightarrow \ket{5, 4}$ & $0.426$ & 7.435 & $19.6$ \\
 &  & $\ket{4,4} \leftrightarrow \ket{5,5}$ & $0.475$ & 7.452 & $25.3$ \\
\hline
\end{tabular} 
  \caption{Relevant Si:Bi transitions and their characteristics for $B_0=3\,$mT. Highlighted in red are the transitions accessible to our resonators}
		\label{tbl:Transitions}
\end{table}
%
%

\section{Experimental details}

Details on the bismuth implanted sample and an extensive description of the setup are included in \cite{QuantumESR}. We present in the following the exact protocols used to acquire the data shown in Figure 2 and 3 of the main text.

\subsection*{Experimental determination of $T_1$ at resonance}

\begin{figure}[!htbp]
  \centering
  \includegraphics[width=180mm]{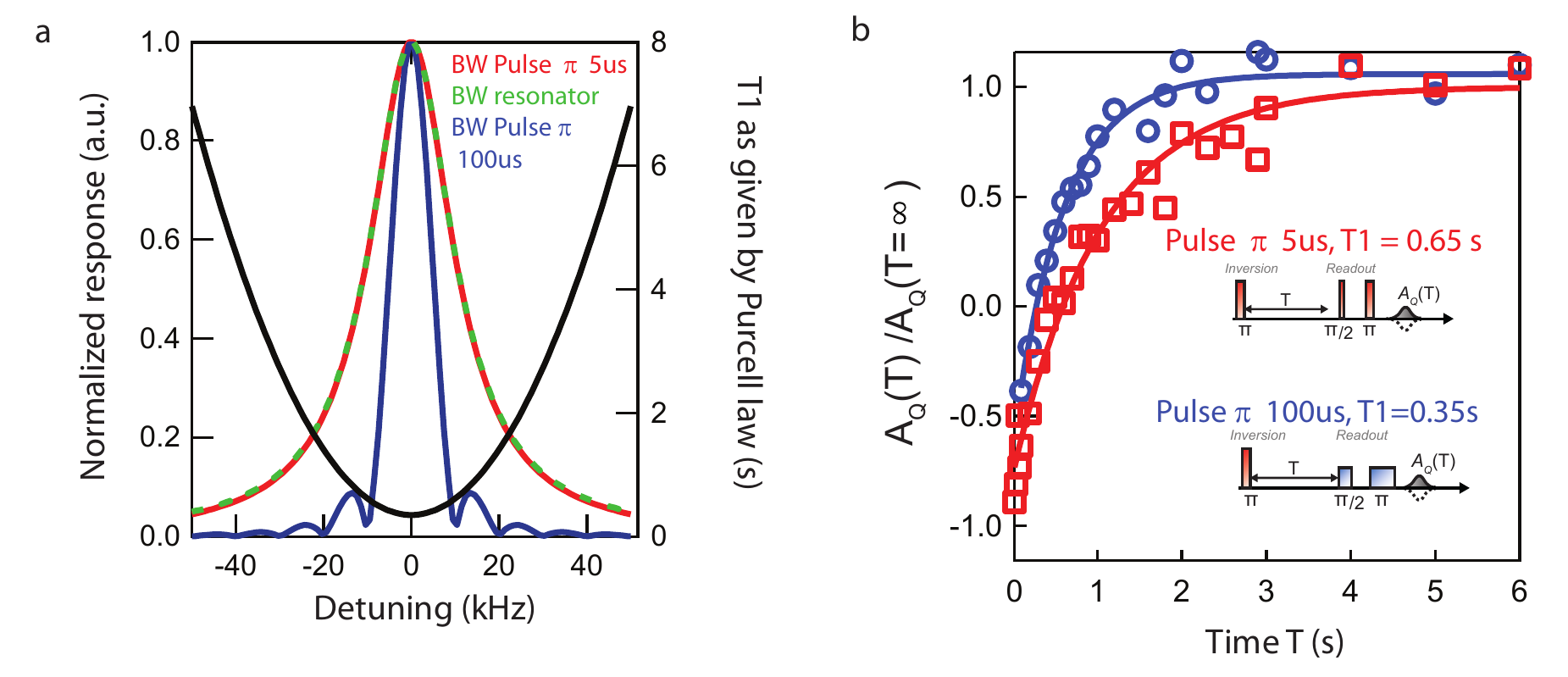}
  \caption{Excitation pulse bandwidth effect on $T_1$ measurement. (a) Computed pulse bandwidth, respectively for a $5(100)$-$\mu$s $\pi$ pulse, in red (blue) incident on a cavity with $\kappa/2\pi=23\,$kHz (green dashes). To illustrate the averaging effect of the pulse bandwidth on $T_{1}$ measurements, the expected Purcell $T_1$ curve (black dashes) as a function of spin-cavity detuning is plotted on the right scale, with $T_1(0)=0.35\,$s and $\kappa/2\pi=23\,$kHz. (b) $T_1$ measurements for two different $\pi$ pulse lengths, measured on resonance with resonator A. $\pi=100\mu\,$s (red) yields $T_1=0.35\,$s, which is in agreement with the Purcell rate . $\pi=5\mu\,$s (red) yields $T_1=0.65\,$s, a factor 2 away from the accurate value.}
		\label{figSuppT1BW}
\end{figure}

This part aims to explain the inversion recovery sequence presented in Figure 2 of the main text. If we rewrite Eq.1 of the main text, we can express $$T_1(\delta) = T_{1}(0) (1+ 4 \delta^2/\kappa^2) \label{eq:T1delta}$$
 with $T_{1}(0)=\kappa/4g^2$. Since the probed ensemble of spins has a larger linewidth $\Delta\omega= 2\,$MHz than our resonators, the signal emitted during the spin-echo comes from a subset of the ensemble of spins, with a frequency spectrum at least as large as the resonator bandwidth. Spins probed at the edges of the bandwidth of the resonator will have longer Purcell relaxation times: for instance those detuned by $\delta=\kappa$ have an expected Purcell relaxation time five times slower than the $T_1$ time expected at perfect resonance, Figure~\ref{figSuppT1BW}a. The contribution of those spins with a longer decay time to the signal will result in an averaging effect, meaning that the measured $T_1$ will be erroneously longer than predicted. 

In order to suppress this effect, we reduce the bandwidth of the readout sequence so as to collect signal only from spins very close to the resonance. The response function of a pulse of length $t_p$ incident on a cavity with bandwidth $\kappa$ at frequency $\omega_0$ is expressed as :
$$ \mathcal{R}(\omega)=2 \frac{ \sin(t_p(\omega-\omega_0)/2)}{t_p(\omega-\omega_0)} \times \mathcal{R}_{cav}(\omega) = 2 \frac{ \sin(t_p(\omega-\omega_0)/2)}{t_p(\omega-\omega_0)} \times \frac{1}{1+4 \left(\frac{\omega-\omega_0}{\kappa}\right)^2}$$

As shown on Figure~\ref{figSuppT1BW}a, for the narrowest bandwidth $\kappa/2\pi=23\,$kHz of resonator A, pulses of $5\upmu\,$s are heavily filtered by the resonator and have the same bandwidth whereas $100\upmu$s-long pulses have a reduced bandwidth of $\approx 10$kHz. In case of $100\upmu$s-long excitation pulses, the Rabi frequency is such that only spins with $\lvert \delta\lvert/2\pi \leq 5\,$kHz will contribute to the signal. This corresponds to a dispersion of only 5\% for the expected Purcell relaxation times, which is negligible. To illustrate the averaging effect, two inversion recovery curves are shown on Figure~\ref{figSuppT1BW}b with readout pulses of $5\upmu$s and $100\upmu$s. The former yields $T_1=0.65\,$s, which is a factor $2$ higher than predicted by the Purcell effect whereas the lattest yields the expected value $T_1=0.35\,$s.

Thus Figure~2d of the main text shows an inversion recovery sequence that has a readout echo sequence with a narrow bandwidth ($t_{\pi }=100\upmu$s, $t_{\pi/2}=t_{\pi }/2$) to suppress contribution from spins with a lower decay rate, and an inversion pulse with large-bandwidth ($t_{\pi}=5\upmu$s) in order to maximize the efficiency of the inversion.

\subsection*{Protocol used to measure spin-cavity detuning dependent relaxation rate}

The goal of this section is to detail the protocol used in Figure 3 of the main text to study the Purcell rate dependence on the spin-cavity detuning, $\delta$.

In order to study this dependence, we detune the spins from the cavity by applying a magnetic field pulse. This is done in our setup by adding a pulse generator with $50 \Omega$ output impedance in parallel to the DC supply of one of the Helmholtz coils which have a $1\,$Hz bandwidth. To minimize the effect of transients, buffer times of 1s are added after ramping the coil up and down. To limit the loss of signal during those buffer times, we use an angle $\theta=45\degree$ and work with resonator B in order to have a longer $T_{1}(0)=1.68\,$s. This $T_1$ was measured with inversion recovery. All the data presented in this section as well as in Figure 3 of the main text were done in a separate run. The quality factor of resonator B dropped from $Q_B=1.07 \times 10^5$ to  $Q_B=8.9 \times 10^4$ due to slightly higher losses, yielding the resonator bandwidth $\kappa/2\pi=82\,$kHz.

\begin{figure}[!htbp]
  \centering
  \includegraphics[width=170mm]{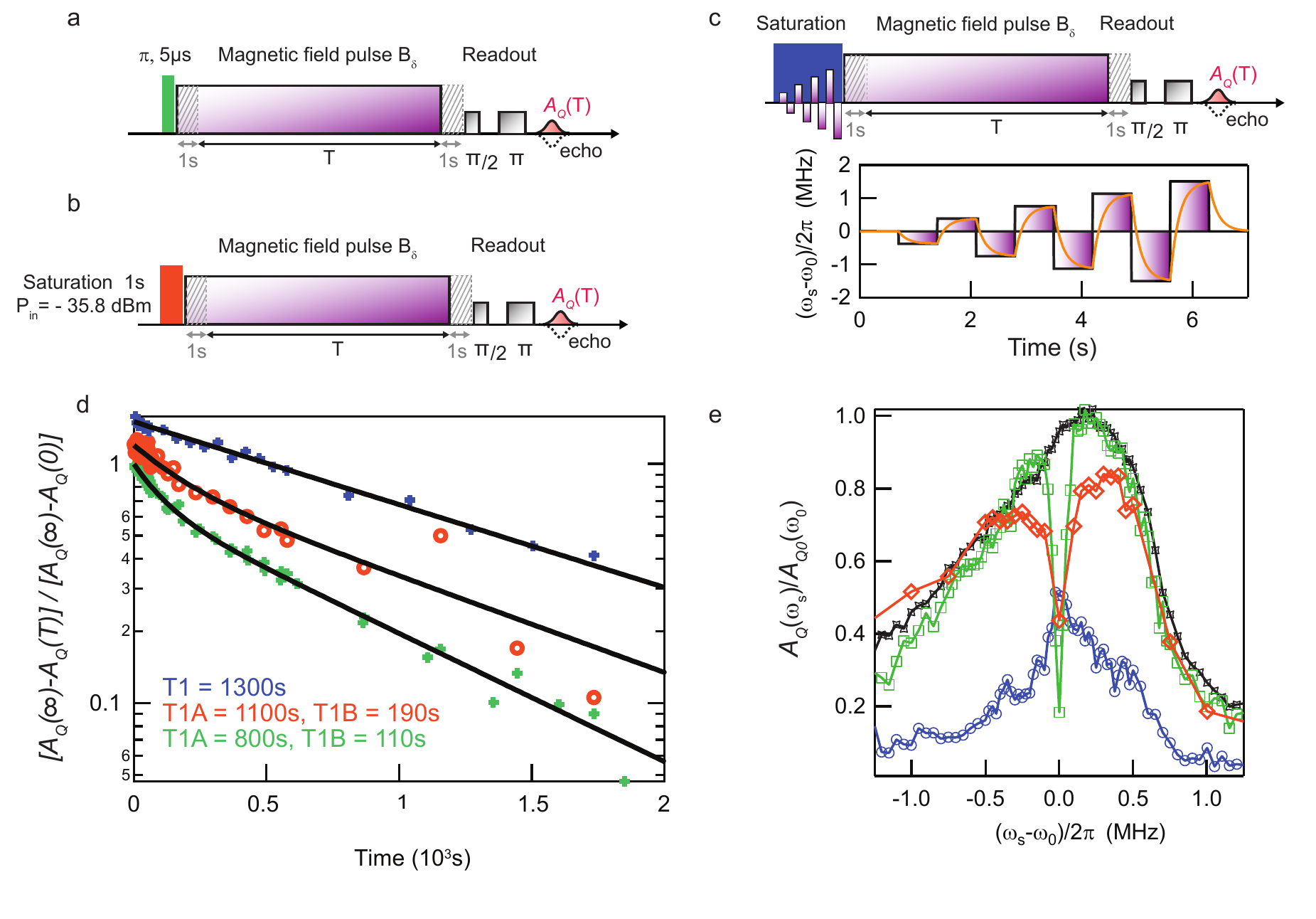}
  \caption{Spectral spin diffusion. (a,b,c) $T_1$ measurement sequence when spins are detuned from the cavity by applying a magnetic field $B_{\delta}$, providing a detuning $\delta=\omega_s-\omega_0=2 \pi \gamma_{\rm eff} B_{\delta} $, with $\gamma_{\rm eff}=df/dB (B_0)$. \textbf{a}) uses a $\pi=5\upmu$s pulse to realise a so-called inversion recovery sequence, \textbf{b}) and \textbf{c}) are saturation recovery sequences: \textbf{b}) uses a 1-s-long strong microwave pulse sent at cavity resonance whereas \textbf{c} has in addition a magnetic field scan shown on the bottom part. Depicted in orange is the idealized magnetic field profile due to the 1-Hz bandwidth of the coil (orange).  (d) $T_1$ measurements for  sequence \textbf{a} (green), \textbf{b} (orange), \textbf{c} (blue) for $\delta=3.8\,$MHz. Fits (black lines): \textbf{a} \& \textbf{b} have a double exponential decay whereas \textbf{c} is a simple exponential. We attribute this double-exponential decay to spin diffusion. (e) Spectral profiles of excitation pulses \textbf{a} (green), \textbf{b} (orange) \& \textbf{c} (blue). The sequence is as follows: send the excitation pulse, detune the spins and measure $A_{Q}(\omega_s)$. Black line is the reference profile without any excitation pulse, yielding reference $\langle S_z(\omega_s) \rangle=-A_{Q0}(\omega_s)/A_{Q0}(\omega_s)$. When an excitation pulse is sent, one can access $\langle S_z(\omega_s) \rangle=-A_{Q}(\omega_s)/A_{Q0}(\omega_s)$.  Note that neither the $\pi$ profile or the saturation profile reach either the full inversion +1 or full saturation 0 at resonance. This is an artefact due to the coil transient time.}
		\label{figSuppT1sat}
\end{figure}

To observe the long relaxation times, such as those measured in Figure~3 of the main text, inversion recovery is not an ideal method. Indeed, when the spin linewidth is broader ($\sim\times 20$) than the excitation bandwidth and when the thermalization time is very long, one can observe polarization mixing mechanisms \cite{Bloembergen1949,Abragam.NuclearMagneticResonance}, spectral and spatial spin diffusion being the most relevant to our case, as the system is only constituted from one species. If one tries to measure the relaxation from spins that have been detuned by an amount $\delta/2\pi=(\omega_s-\omega_0)/2\pi= 3.8\,$MHz during a lapse of time $T$ with an inversion recovery sequence (Figure~\ref{figSuppT1sat}a), one observes a double exponential relaxation (Figure~\ref{figSuppT1sat}d, green), pointing towards the existence of a spin diffusion mechanism.

Spin diffusion is prevented by suppressing any polarization gradient along the spin line, which leads us to use a saturation recovery scheme instead of inversion recovery. The simplest saturation recovery scheme (Figure~\ref{figSuppT1sat}b) consists of sending a strong microwave tone resulting in the saturation of the line, producing an incoherent mixed state with the population evenly shared between excited and ground states. Nevertheless, a relaxation time measured with this scheme still yields a double-exponential decay (Figure~\ref{figSuppT1sat}d, orange), with time constant similar to the inversion recovery case. This implies that the saturation of the line was insufficient.

To improve the saturation, one can sweep the magnetic field during the saturation pulse so as to bring different subsets of the spin line to resonance and realize a full saturation. The adopted sweep scheme is shown on Figure~\ref{figSuppT1sat}c. The relaxation curve acquired with such a curve is a simple exponential (Figure~\ref{figSuppT1sat}d, blue), indicating the suppression of the spin diffusion effect.

One can further check the quality of the saturation by measuring the polarization across the full spin linewidth immediately after saturation. To realize such scans (Figure~\ref{figSuppT1sat}e), we apply the relevant saturation pulse at $\omega_0$, then apply a magnetic field pulse $B_{\delta}=(\omega_s-\omega_0)/\gamma_e$ and measure the echo signal $A_Q(\omega_s)$ with a Hahn echo sequence. When no saturation pulse is applied, the measured echo signal $A_{Q0}(\omega_s)$ is a measure of the full polarization $-\langle S_z(\omega_s) \rangle = +1$ (black curve) and shows the natural spin linewidth. When studying an excitation pulse, the polarization of the spins is given by $-\langle S_z(\omega_s) \rangle = A_{Q}(\omega_s)/A_{Q0}(\omega_s)$, where $A_{Q}(\omega_s)$ is the measured echo signal. Thus $-\langle S_z(\omega_s) \rangle = -1$  ndicates full inversion, $\langle S_z(\omega_s) \rangle = 0$ saturation and $-\langle S_z(\omega_s) \rangle = +1$ return to thermal equilibrium. The green, orange and blue curves are taken after respectively a $\pi$ pulse (\textbf{a}) and a saturation without field sweeps (\textbf{b}) and with field sweeps (\textbf{c}). At resonance, one expects a change of $S_z$ from -1 to +1  for a $\pi$ pulse and from -1 to 0 for a saturation pulse. Due to the coil transient time, all three curves shows a partial relaxation. If the saturation was optimal and no partial relaxation was occurring, one should observe $S_z=0$ for any detuning $\delta$. Among the two saturations (\textbf{b}) and (\textbf{c}) studied here, only the last saturation scheme (\textbf{c}) equally saturates the line. The basic saturation has a bandwidth $\approx 250\,$kHz and the $\pi$ pulse bandwidth is similar to the cavity $\kappa=82\,$kHz. This confirms that only in scheme (\textbf{c}) can spin diffusion be fully suppressed and yield a simple exponential decay relaxation. This is this last scheme that is used to measure the $22$ relaxation rates at different detunings $\delta$ of Figure~3 of the main text.

The global fit shown on Figure~3c of the main text is obtained by using equation  $T_1(\delta)^{-1}= \Gamma_{\rm P}+\Gamma_{\rm NR}$ which may be expressed as $T_{1}(0)^{-1} \left( 1 + 4 \left( \frac{\delta}{\kappa}\right)^2\right)^{-1} +\Gamma_{\rm NR}$ to involve only experimentally determined parameters. Indeed, $\kappa$ is precisely determined by measuring the quality factor of the resonator at very low power while $T_{1}(0)$ is determined by an inversion recovery sequence as mentioned above. $\delta$ has been determined via precise calibration of the coil pulse, thus the only remaining free parameter in the fit is $\Gamma_{\rm NR}$, yielding  $\Gamma_{NR}^{-1}=1600\,$s. The errors bars come from the accuracy of the relaxation rates fits.

\clearpage

\end{document}